\documentclass[conference]{IEEEtran}
\usepackage{amsmath,amsfonts,amssymb}
\usepackage{graphicx}
\usepackage{caption}
\usepackage{subcaption}
\usepackage{booktabs}
\usepackage{multirow}
\usepackage{url}
\usepackage{cite}
\usepackage{float}
\usepackage{stfloats}
\usepackage{placeins}
\usepackage{multicol}

\title{Dynamic Windowing in Transformers via Regime Incorporation for Financial Time Series}

\author{
\IEEEauthorblockN{Praveen, Prince Chouhan, Keshav Maheshwari, Aman Verma}
\IEEEauthorblockN{22b3931@iitb.ac.in, 22b3970@iitb.ac.in, keshav.maheshwari@iitb.ac.in, 22b3929@iitb.ac.in}
\IEEEauthorblockA{Department of Electrical Engineering\\
Indian Institute of Technology Bombay}
}

\makeatletter
\def\@IEEEpubidpullup{8\baselineskip}
\makeatother

\IEEEpubid{\begin{minipage}{\textwidth}\centering
\textsuperscript{*}These authors contributed equally to this work.
\end{minipage}}

\begin{document}
\maketitle
% \begin{multicols}{2}
\begin{abstract}
Financial time series exhibit non-stationary behavior, where the strength and extent of temporal dependencies change across market regimes. Trending low-volatility phases typically require long-range contextual information, whereas mean-reverting high-volatility periods rely more heavily on short-term dynamics. Standard Transformer architectures, with fixed attention windows and static positional encodings, are therefore unable to adapt to such variations. In this work, we propose a regime-aware dynamic windowing framework that incorporates market regime information directly into the Transformer. We construct four generic regime signals from the price series---volatility ratio, trend strength, local predictability ratio (LPR), and rolling autocorrelation---and embed them into the model through two mechanisms: (i) regime-augmented inputs to standered architecture, and (ii) a modified attention layer that modulates the attention weights using regime embeddings. Experiments on 5 S\&P 500 stocks show consistent improvements across five evaluation metrics, demonstrating that regime-aware dynamic windowing enhances both explanibility and predictive performance in financial forecasting tasks.
\end{abstract}
\section{Introduction}

Financial time series are inherently non-stationary and evolve through a variety of market regimes, each characterized by distinct temporal structures, volatility profiles, and predictability patterns. Classical econometric approaches such as ARIMA and GARCH assume stationary dynamics or predictable variance structures, making them less effective under rapid regime transitions~\cite{hamilton_regime_switching, bollerslev_garch}. Modern deep learning models, especially Transformers, have recently demonstrated strong performance in sequential modeling due to their ability to capture complex dependencies~\cite{vaswani_attention, zerveas_tft}. However, standard Transformer architectures rely on \emph{fixed} positional encodings and global attention with uniform weighting rules, implicitly assuming that the underlying dependency horizon remains constant across time. This assumption rarely holds in real-world financial markets.

Empirical evidence shows that trending phases---typically associated with low volatility---exhibit persistent, long-range temporal dependencies~\cite{lux_trends, lo_persistence}. During such periods, information from distant timesteps remains relevant because the price process displays momentum-driven continuity. Conversely, high-volatility mean-reverting environments exhibit rapid corrections where the predictive structure is dominated by short-term fluctuations, making long-range information less relevant or even detrimental~\cite{cont_stylized_facts}. Hence, the effective temporal footprint required for accurate forecasting is regime-dependent and changes over time.

Despite their global receptive field, practical Transformer-based forecasting models typically impose explicit or implicit window constraints to manage computational cost and noise sensitivity~\cite{lim_tft}. A fixed attention span inevitably becomes suboptimal when temporal dependencies vary. Several works have attempted to address this through hierarchical attention~\cite{liu_logformer}, learned temporal compression~\cite{zhou_informer}, or adaptive positional encoding~\cite{ke_rel_pos_encoding}. However, none of these explicitly incorporate financial regime information or adjust attention spans based on evolving market states.

To address this limitation, we propose a regime-aware dynamic windowing mechanism for Transformer-based financial forecasting. Rather than relying on externally labeled regimes or manually defined market phases, we introduce four \emph{generic, self-supervised regime descriptors} derived directly from raw price series: (i) short-to-long volatility ratio, (ii) trend strength computed from multi-scale moving averages, (iii) local predictability ratio (LPR) comparing short-term and long-term deviations, and (iv) rolling lag-1 autocorrelation capturing temporal persistence. These descriptors are motivated by well-established findings in market microstructure and volatility modeling~\cite{engle_volatility, tsay_book}.

We integrate these regime descriptors into the Transformer framework using two architectural pathways. First, we augment positional encodings with regime embeddings, enabling the model to condition temporal encoding on the prevailing market state. Second, we introduce a regime-modulated attention mechanism where attention weights are adaptively scaled based on regime embeddings, effectively expanding or contracting the receptive field depending on whether the model is in a trending or mean-reverting phase. This approach is lightweight, architecture-agnostic, and does not require explicit segmentation or supervised regime labels, making it practical for real-world deployment.

\subsection{Key Contributions}

\begin{itemize}
    \item We introduce four generic, interpretable regime descriptors derived directly from financial price series, grounded in empirical market structure literature.
    \item We design a regime-aware dynamic windowing Transformer that adaptively adjusts its attention span via regime-signals inputs.
\end{itemize}

\section{Motivation and Background}

Financial time series are well known to exhibit regime-dependent temporal structure, where 
market behaviour alternates between trending low-volatility phases and turbulent mean-reverting 
high-volatility phases~\cite{ang2002regimes}. 
These regimes differ not only in statistical properties such as volatility and autocorrelation, 
but also in the \emph{effective temporal dependencies} relevant for forecasting. 
During smooth trending periods, long-range dependencies dominate and models require a larger 
effective receptive field. Conversely, during volatile or shock-driven regimes, the dynamics are 
short-lived and dominated by rapid corrections, making short-range patterns more informative. 
Traditional forecasting methods---ARIMA, GARCH-family models, and Markov-switching models---treat 
regimes as latent discrete states and employ separate model parameters per regime, 
with relatively rigid transition rules. Such hard-switching approaches often 
struggle to generalize to the fine-grained, continuously evolving regimes that modern markets exhibit.

Transformers have emerged as competitive sequence models for financial forecasting due to their 
ability to capture long-range dependencies through multi-head self-attention~\cite{vaswani_attention}. 
However, standard Transformers assume \emph{static} positional encodings and a \emph{fixed} attention 
span across the entire time series. This assumption is fundamentally incompatible with financial 
regimes, where the useful context window can vary drastically over time. Recent studies have shown 
that attention maps extracted from pretrained models exhibit sharp regime-specific behaviour: 
in trending zones, heads attend farther into history, whereas in high-volatility periods, attention 
contracts near the recent past~\cite{zerveas2021ts_transformer}. This motivates the need 
for \emph{regime-aware dynamic windowing} in attention-based architectures.

A second source of complexity arises from feature representation. Modern forecasting systems often 
combine time-domain features (lags, realized volatility, returns, microstructure variables) with 
frequency-domain features computed using STFT, wavelets, or fractional Fourier transforms. Time-domain 
features excel at capturing local dynamics, while frequency-domain features capture periodicities, 
cycles, and structural oscillations. Concatenating both is a common and strong baseline due to 
its simplicity and robustness. However, this approach suffers from several limitations:
(1) the chosen temporal and spectral resolutions are globally fixed and do not adapt across regimes;
(2) redundant or noisy coefficients inflate dimensionality and can amplify overfitting; and
(3) the model must implicitly learn when to trust time-domain vs. frequency-domain features, which 
requires larger datasets and imposes unnecessary burden on the learning process. 

These challenges highlight the need for a mechanism that not only incorporates regime information 
but also allows regimes to \emph{control} the model’s effective receptive field and the 
relative weighting of heterogeneous features. Simply appending regime indicators as features 
provides no guarantee that the model will learn the desired modulation; instead, the signal becomes 
entangled nonlinearly with the rest of the input, losing interpretability and degrading robustness. 
In contrast, regime-aware architectures inspired by mixture-of-experts~\cite{shazeer2017moe}, 
context gating~\cite{pascanu2013gating}, or adaptive attention mechanisms provide explicit 
control over feature importance.

Motivated by this, we design a dynamic-window Transformer architecture that integrates regime 
information directly into the attention mechanism. We propose two variants:
(i) \textbf{Architecture~0}, which augments the standard Transformer by adding 
regime embeddings to the token representations before feeding them to the self-attention layers; and 
(ii) \textbf{Architecture~1}, which modifies the attention block itself by 
\emph{masking or modulating} the vanilla Transformer’s attention weights using regime embeddings, 
thereby enabling dynamic expansion and contraction of attention span.  
This formulation aligns naturally with the behaviour observed in financial markets, where the 
relevant temporal horizon and feature relevance shift continuously with underlying regimes.

Together, these ideas form a principled framework for regime-aware forecasting that bridges 
traditional econometric insights on regime-switching with modern deep sequence models, while 
explicitly addressing the time–frequency tradeoff and dynamic dependency structure inherent 
to financial data.

\section{Methodology}
In this section, we outline the overall methodology used in our study. We first describe the baseline Transformer model employed for financial time series forecasting. We then introduce our two proposed extensions that incorporate regime information: Architecture~0, which appends regime embeddings to the input representations, and Architecture~1, which integrates regime signals directly into the attention mechanism by modulating the attention mask. Central to both designs are four regime features—volatility ratio, trend strength, LPR, and rolling lag-1 autocorrelation—extracted directly from the price series to characterize market conditions. These components collectively form a flexible framework for enabling dynamic windowing based on underlying market regimes.

\subsection{Regime Time Series Formulation}

Let $p_t$ denote the asset price and $r_t = \log(p_t/p_{t-1})$ denote log-returns.  
We construct four generic regime signals designed to capture different structural properties of financial time series.  
Each signal is bounded using a $\tanh(\cdot)$ transform to stabilize extreme values and improve training dynamics.  

All regime–feature plots shown in the next four subsections are computed using **BTC-USD daily data from 2018 to 2023**.  
For reference, the corresponding **daily closing price of BTC-USD** over this period is provided below in Figure~\ref{fig:btc_price}.

\vspace{1em}

\begin{figure}[H]
    \centering
    \includegraphics[width=0.92\linewidth]{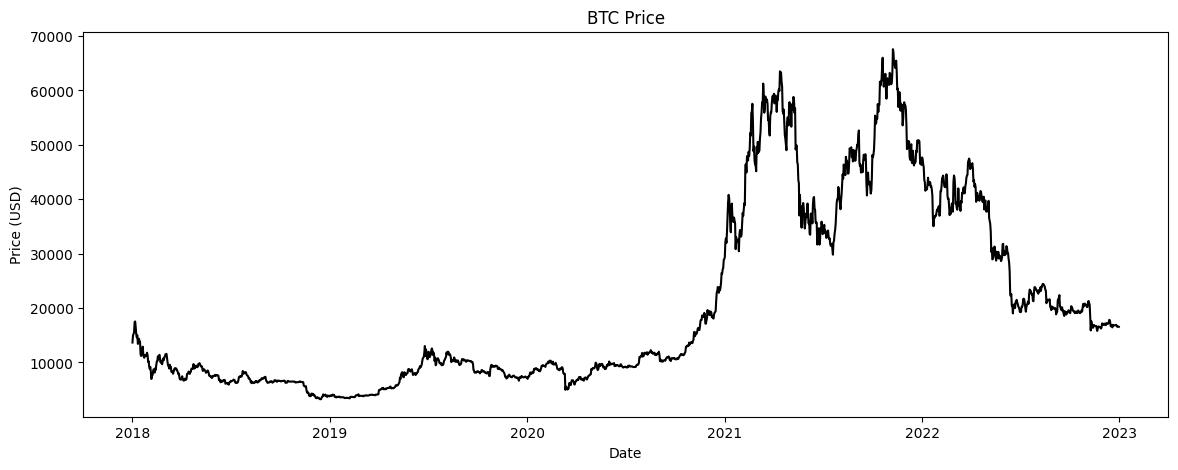}
    \caption{Daily closing price of BTC-USD from 2018 to 2023}
    \label{fig:btc_price}
\end{figure}

\vspace{1em}

1) \textbf{Volatility Ratio (VR).}  
This measures the relative intensity of short-term volatility compared to long-term volatility:
\begin{equation}
\text{VR}_t = \tanh\left(\frac{\sigma_{s}(t)}{\sigma_{l}(t) + 10^{-8}}\right),
\end{equation}
where $\sigma_s$ and $\sigma_l$ denote rolling standard deviations over short and long windows.  
A rising volatility ratio typically corresponds to market stress or rapid regime shifts.

\begin{figure}[H]
    \centering
    \includegraphics[width=0.92\linewidth]{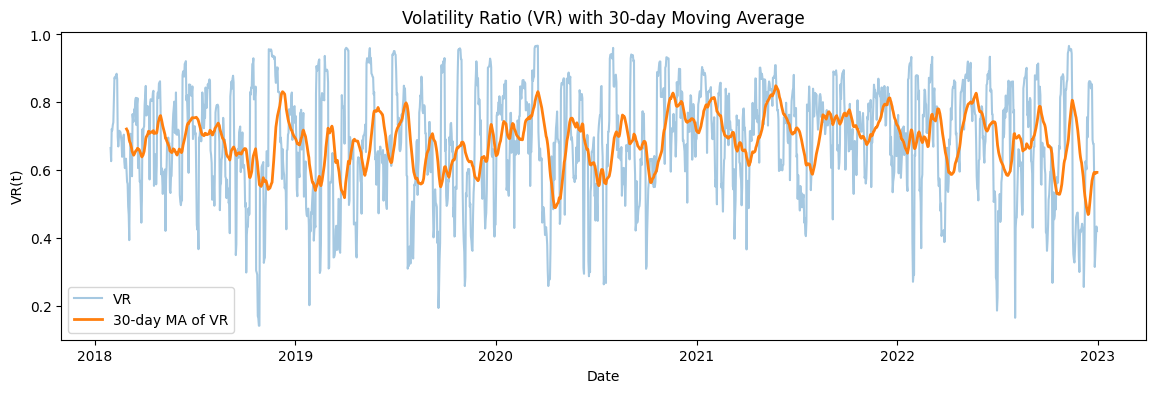}
    \caption{Volatility Ratio (VR) for BTC-USD}
\end{figure}

2) \textbf{Trend Strength (TS).}  
This captures the magnitude of the deviation between long- and short-term moving averages:
\begin{equation}
\text{TS}_t = \tanh\left(\frac{|MA_l(t)-MA_s(t)|}{|p_t| + 10^{-8}}\right),
\end{equation}
where $MA_s$ and $MA_l$ denote short- and long-term moving averages.  
Higher values indicate stronger directional trends or persistent drift in price.  

\begin{figure}[H]
    \centering
    \includegraphics[width=0.92\linewidth]{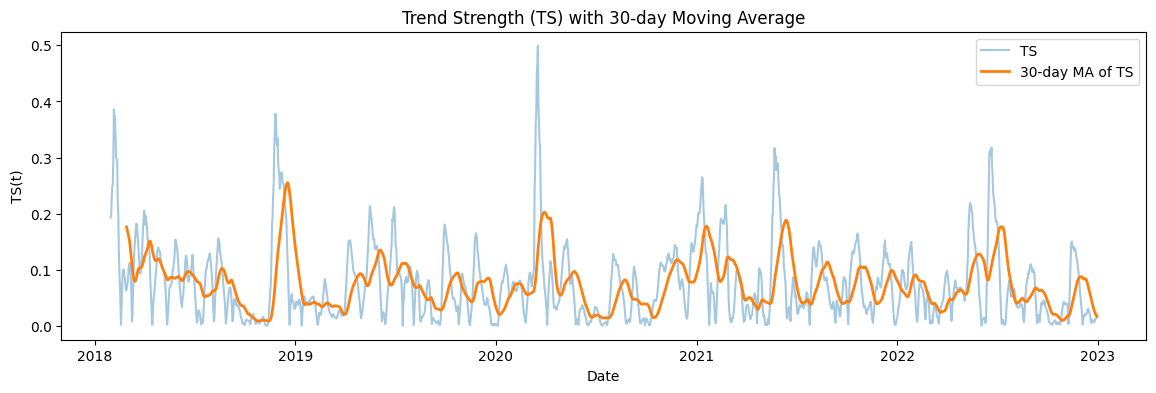}
    \caption{Trend Strength (TS) for BTC-USD}
\end{figure}

3) \textbf{Local Predictability Ratio (LPR).}  
This compares short-term deviation errors against long-term deviation errors:
\begin{equation}
\text{LPR}_t = \tanh\left(\frac{E_l(t)-E_s(t)}{E_l(t)+E_s(t)+10^{-8}}\right),
\end{equation}
where $E_s(t)=|p_t - MA_s(t)|$ and $E_l(t)=|p_t - MA_l(t)|$.  
Positive values imply that recent prices align more closely with short-horizon structure, indicating locally predictable behaviour.  

\begin{figure}[H]
    \centering
    \includegraphics[width=0.92\linewidth]{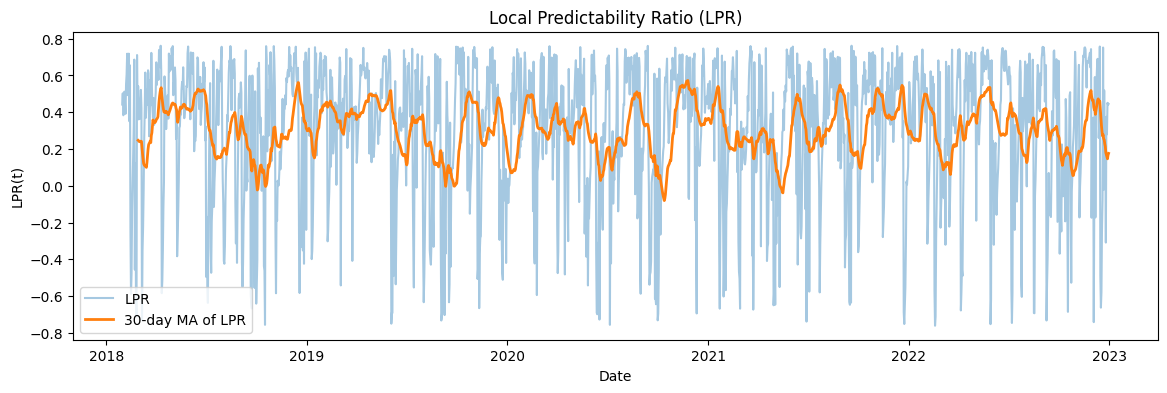}
    \caption{Local Predictability Ratio (LPR) for BTC-USD}
\end{figure}

4) \textbf{Rolling Lag-1 Autocorrelation ($\rho_t$).}  
This quantifies the persistence or mean reversion in recent returns:

\begin{equation}
\rho_t = \text{ACF}(r_{t-w:t}, \text{lag}=1).
\end{equation}

Positive autocorrelation suggests momentum-driven regimes, while negative values indicate mean-reversion or choppy markets.  

\begin{figure}[H]
    \centering
    \includegraphics[width=0.92\linewidth]{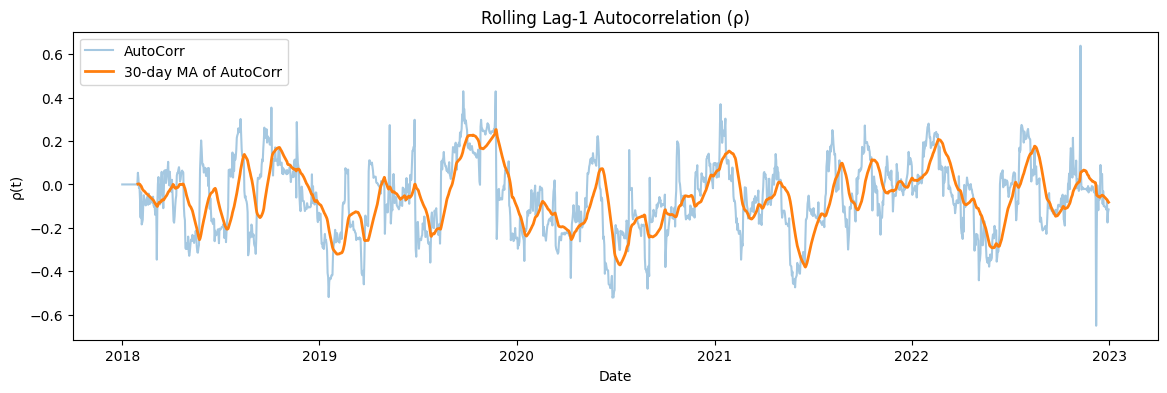}
    \caption{Rolling Lag-1 Autocorrelation for BTC-USD}
\end{figure}

\subsection{Baseline Transformer Architecture}

We begin with a standard encoder--decoder Transformer operating on a fixed
sequence length of $T = 60$ tokens.  
Each input window of 60 historical prices is first embedded into a latent space
and passed through the encoder, which consists of stacked multi-head
self-attention layers followed by position-wise feedforward transformations.
These layers allow the encoder to extract temporal dependencies across the full
input window.

The decoder receives a start token and autoregressively generates the next price
value. It applies masked self-attention to avoid peeking into the future, and
cross-attention over the encoder outputs to utilize information from the input
sequence.  
Together, this encoder--decoder design naturally models conditional forecasting
tasks with flexible dependency structures.

The core of the Transformer is the scaled dot-product attention:
\begin{equation}
\mathrm{Attn}(Q,K,V)
= \mathrm{softmax}\left( \frac{QK^T}{\sqrt{d_k}} \right) V ,
\end{equation}
where $Q = XW_Q$, $K = XW_K$, and $V = XW_V$ arise from learned projections of
the input embeddings.  
This formulation defines the ``baseline'' attention mechanism against which our
regime-aware extensions are compared.

\vspace{1em}

\begin{figure}[H]
    \centering
    \includegraphics[width=0.95\linewidth]{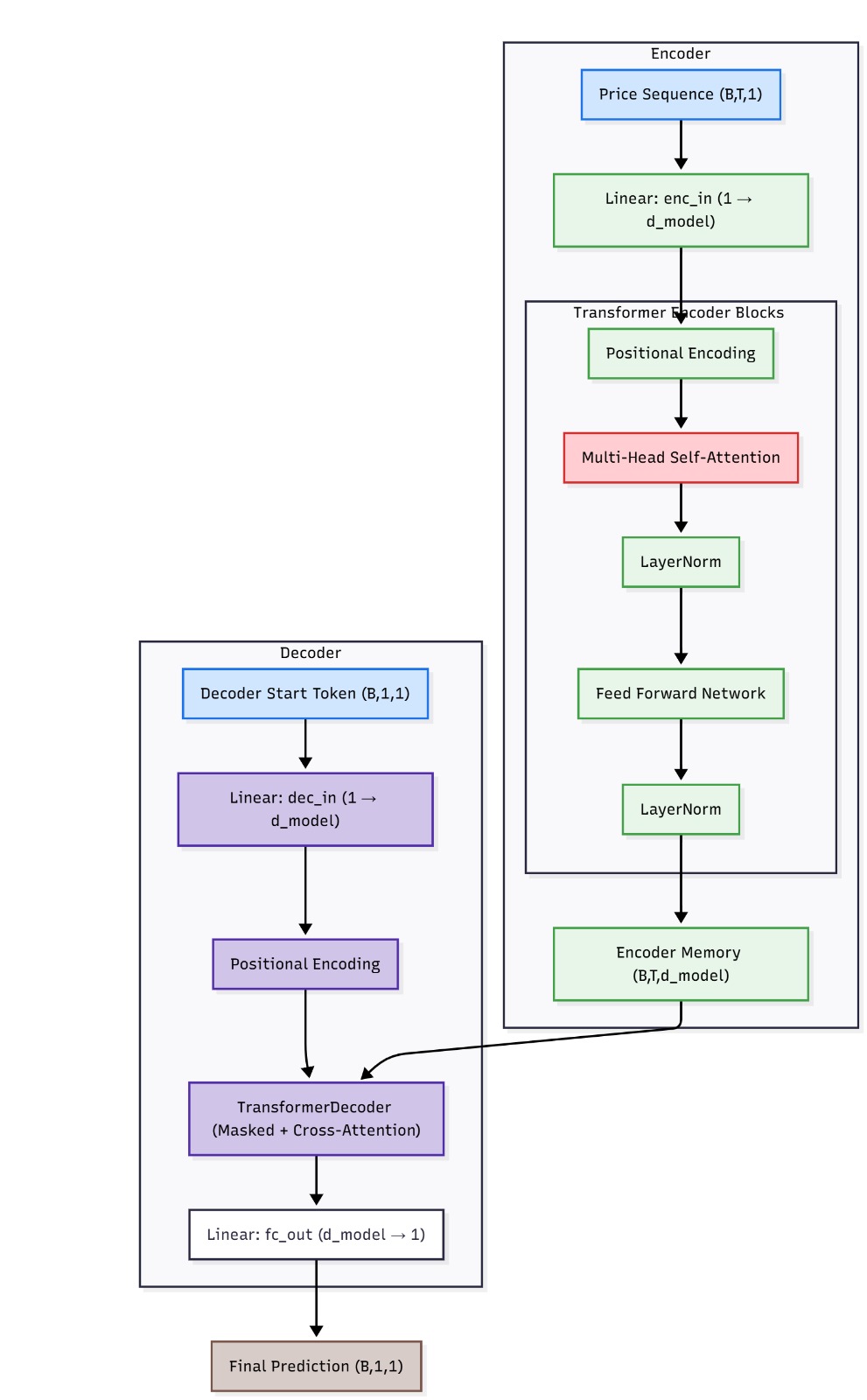}
    \caption{Baseline encoder--decoder Transformer architecture with sequence length 60.}
    \label{fig:baseline_transformer}
\end{figure}

% ============================================================
\subsection{Architecture 0: Regime-Augmented Positional Encoding}

In this architecture, regime information is injected directly at the input-level
by modifying each token representation.  
The intuition is that regime dynamics—such as volatility, trend strength,
predictability, and autocorrelation—represent exogenous signals that influence
how the model should interpret temporal patterns.  
By enriching token embeddings with regime descriptors, the Transformer receives
a context-aware representation before any self-attention operations occur.

Given the token embedding $X_t$, positional encoding $PE_t$, and regime
embedding $e_t^R$, the combined representation becomes:
\begin{equation}
X'_t = X_t + PE_t + e^R_t .
\end{equation}
Because the Transformer is permutation-invariant, positional encodings ensure
temporal ordering, while regime embeddings allow the model to distinguish
between structurally different market environments.  
This modifies the input without altering the attention mechanism itself.

\vspace{1em}

\begin{figure}[H]
    \centering
    \includegraphics[width=0.95\linewidth]{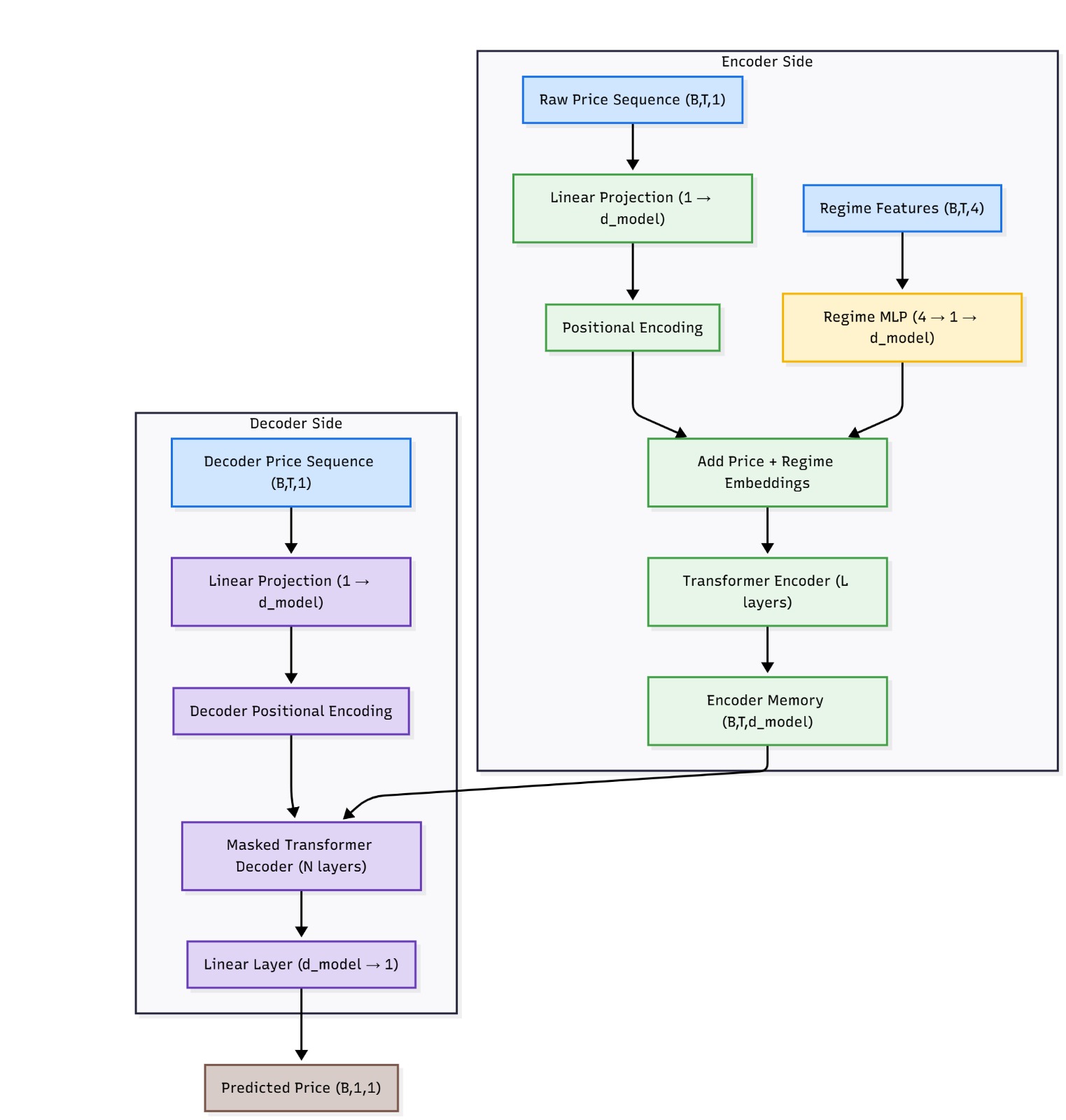}
    \caption{Architecture 0: Regime embeddings and positional encoding added to the input.}
    \label{fig:arch0}
\end{figure}

% ============================================================
\subsection{Architecture 1: Regime-Modulated Attention}
The motivation behind this modification is to allow the model to dynamically
reshape its effective attention window depending on the current market regime.
In standard self-attention, the term $QK^T$ encodes pairwise similarity between
tokens and determines which past positions each query attends to. This produces
a fixed inductive behavior that does not explicitly adapt to changing market
conditions.

By adding a regime-derived bias term $W_r R^T$ directly inside the attention
logits, we allow the model to alter these affinities in a context-dependent
manner. Concretely, if the regime embedding indicates a high-volatility period,
the learned projection $W_r$ may produce a bias that suppresses long-range
attentions and strengthens short-range interactions, effectively shrinking the
active receptive field. Conversely, during stable or strongly trending regimes,
the bias may encourage broader temporal dependencies by amplifying attention to
distant positions. The regime term acts elementwise on the pre-softmax attention matrix,
analogous to a learned, input-dependent modification of the usual attention
mask. Unlike a static triangular or causal mask, this regime-conditioned term
can selectively boost or dampen specific rows or columns of the attention
logits, thereby expanding or contracting the effective attention window on
different time segments. In this way, the model can make its temporal
reasoning adaptive to market structure rather than fixed across all contexts.

\begin{equation}
O =
\mathrm{softmax}\left(
\frac{QK^{T} + QR^{T}}
{\sqrt{d_k}}
\right) V ,
\label{eq:regime_attention}
\end{equation}
where the term $W_r R^{T}$ serves as a regime-dependent adjustment to the
standard $QK^T$ affinity matrix.  
In effect, the regime operates as a dynamic mask that shifts attention weights
toward or away from different temporal positions.

\vspace{1em}

\begin{figure}[H]
    \centering
    \includegraphics[width=0.92\linewidth]{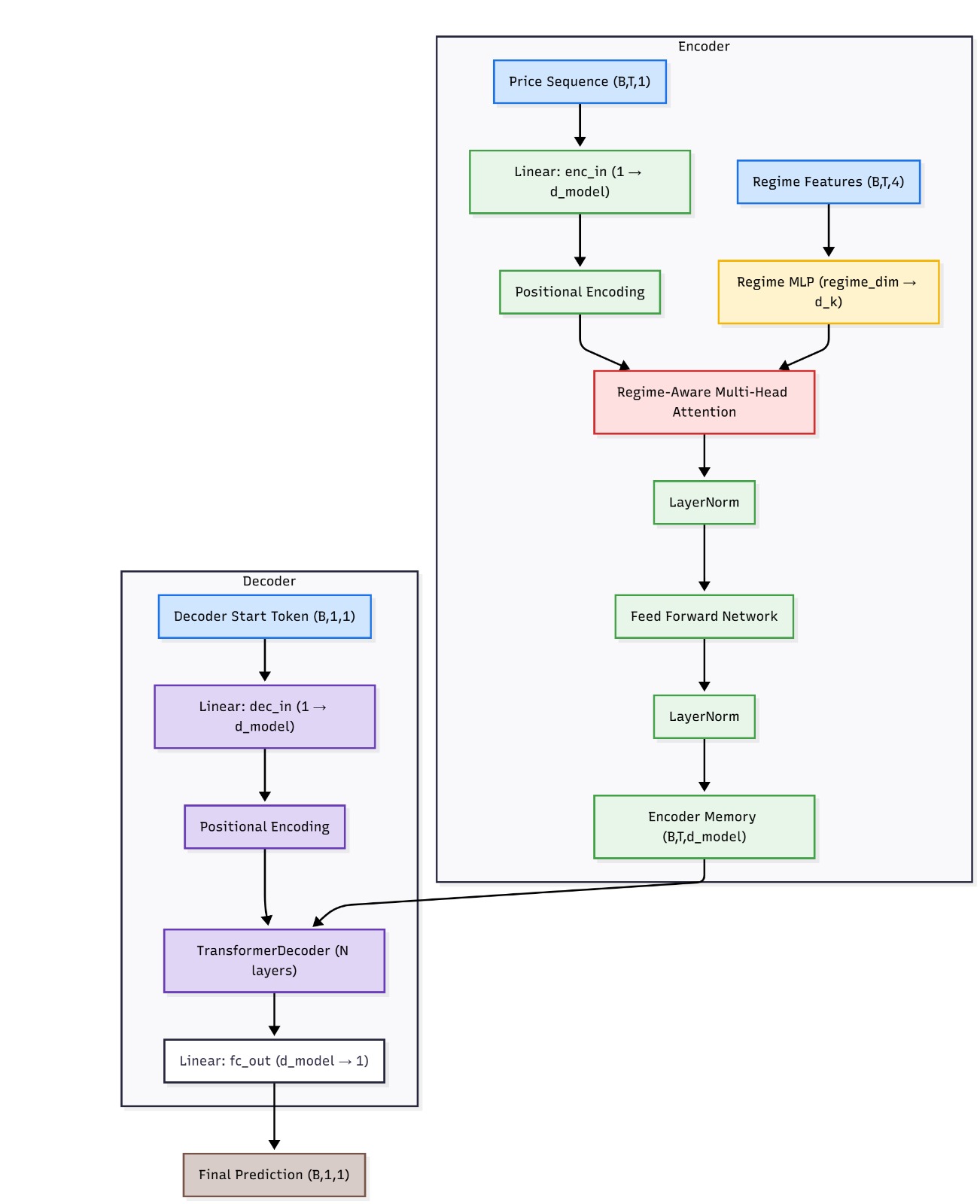}
    \caption{Architecture 1: Regime-modulated Transformer architecture.}
    \label{fig:arch1}
\end{figure}

\begin{figure}[H]
    \centering
    \includegraphics[width=0.92\linewidth]{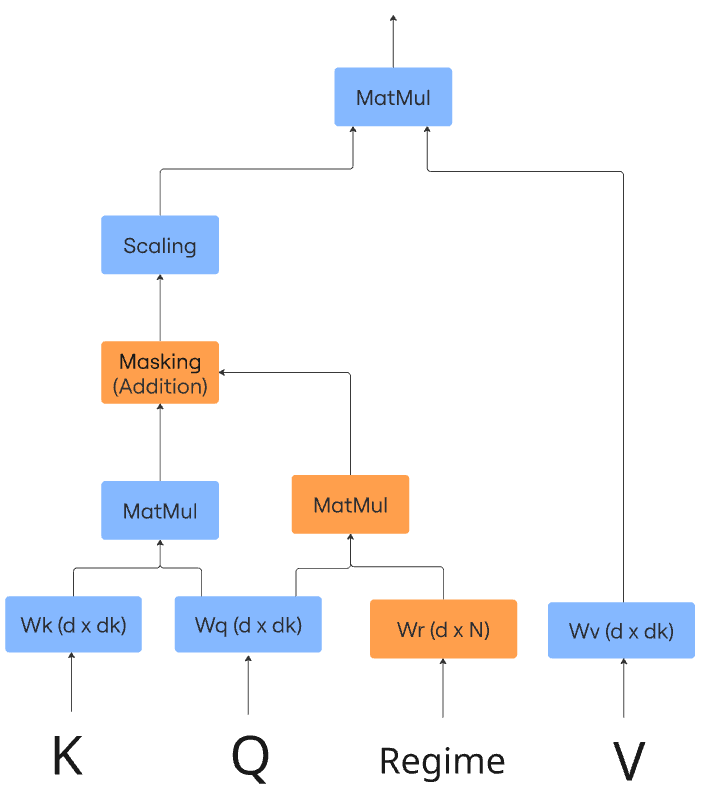}
    \caption{Architecture 1: Regime-modulated attention block with additive regime bias.}
    \label{fig:arch1_attn}
\end{figure}

\section{Results}

\subsection{Evaluation Protocol: Rolling-Origin Backtesting}

To ensure realistic, forward-looking evaluation, we adopt a
\textbf{rolling-origin expanding-window} backtesting procedure.
For each asset, the historical price series is divided into
multiple folds, where the training set expands forward in time while
the test set moves as a non-overlapping future segment.  
This prevents temporal leakage and mimics how forecasting models would
be deployed in practice.

We use the following rolling folds:
\[
(0.10, 0.15),\,
(0.15, 0.20),\,
\dots,\,
(0.85, 0.90),\,
(0.90, 0.95),
\]
where each pair $(a,b)$ denotes training on the first $a$ fraction of
the data and testing on the next $b-a$ fraction.  
Across all folds, predictions are concatenated to compute global metrics.

This method enables model comparison under multiple market regimes,  
including trending periods, sideways ranges, high-volatility phases,  
and sudden reversals, which is critical for evaluating regime-aware architectures.

\vspace{1em}

\subsection{Metrics}

We evaluate all models using five widely used predictive performance metrics:

\begin{itemize}
    \item \textbf{Directional Accuracy (DA)}:  
    Measures the percentage of times the model correctly predicts the direction
    (up or down) of the next-day return.  
    It captures a model's trading relevance rather than raw error magnitude.

    \item \textbf{Mean Squared Error (MSE)}:  
    Penalizes large errors more heavily and measures overall forecasting accuracy.

    \item \textbf{Mean Absolute Percentage Error (MAPE)}:  
    Scale-normalized forecasting error measuring proportional deviation.

    \item \textbf{Pearson Correlation}:  
    Linear correlation between predicted and actual prices.  
    Higher values indicate stronger linear agreement.

    \item \textbf{Spearman Correlation}:  
    Rank-based monotonic relationship between predicted and actual series.  
    Useful for assessing models that capture ordinal trends rather than magnitudes.
\end{itemize}

Together, these metrics allow us to evaluate not only raw numerical accuracy
(MSE, MAPE) but also forecasting structure (correlations) and trading value
(DA).

\vspace{1em}

\subsection{Performance Summary}

Table~\ref{tab:results_all} (reproduced below) reports results for 
\textbf{five S\&P~500 tickers} across the three architectures:
\begin{enumerate}
    \item Baseline Transformer  
    \item Architecture~0 (Regime-Augmented Positional Encoding)  
    \item Architecture~1 (Regime-Modulated Attention)
\end{enumerate}

\vspace{0.5em}

\begin{center}
\textit{A full-page comparison table is included below.}
\end{center}

\begin{table}[H]
\centering
\caption{Performance comparison across three architectures and five stocks over four metrics.}
\label{tab:results_all}
\renewcommand{\arraystretch}{1.25}
\begin{tabular}{|l|c c c c|c c c c|c c c c|}
\hline
\multirow{2}{*}{Stock}
& \multicolumn{4}{c|}{Baseline}
& \multicolumn{4}{c|}{Architecture 0}
& \multicolumn{4}{c|}{Architecture 1} \\
\cline{2-13}
& DA & RMSE & MAPE & Pearson
& DA & RMSE & MAPE & Pearson
& DA & RMSE & MAPE & Pearson \\
\hline

AAPL
& \textbf{0.5034} & 8.3701 & 9.5960 & \textbf{0.9889}
& \textbf{0.5027} & 6.7371 & 7.2604 & 0.9881
& 0.4871 & 7.0646 & 7.9126 & \textbf{0.9893} \\
\hline

TSLA
& 0.4979 & 36.9139 & 14.5754 & 0.9145
& 0.4940 & 31.8779 & 12.3068 & 0.9365
& 0.4933 & 32.1910 & 13.7132 & 0.9358 \\
\hline

BTC-USD
& 0.4857 & \textbf{8976.3273} & \textbf{21.0792} & 0.8558
& 0.4838 & \textbf{7495.5855} & \textbf{18.0996} & \textbf{0.8998}
& \textbf{0.4890} & \textbf{7547.6099} & \textbf{17.9437} & 0.8971 \\
\hline

GOOGL
& 0.5003 & 7.3101 & 5.5814 & 0.9770
& \textbf{0.5115} & 6.7282 & 5.2584 & \textbf{0.9823}
& 0.5027 & 6.6172 & 5.5302 & 0.9816 \\
\hline

NVDA
& 0.4854 & 1.5879 & 11.8035 & 0.9773
& 0.4857 & 1.2612 & 9.5765 & 0.9865
& 0.4891 & 1.7430 & 12.1752 & 0.9807 \\
\hline

\end{tabular}
\end{table}

\FloatBarrier  % ===== Prevent figures from jumping above the table =====

% ==========================================================
% ======================   FIGURES   =======================
% ==========================================================

\begin{figure}[H]
    \centering
    \includegraphics[width=0.95\linewidth]{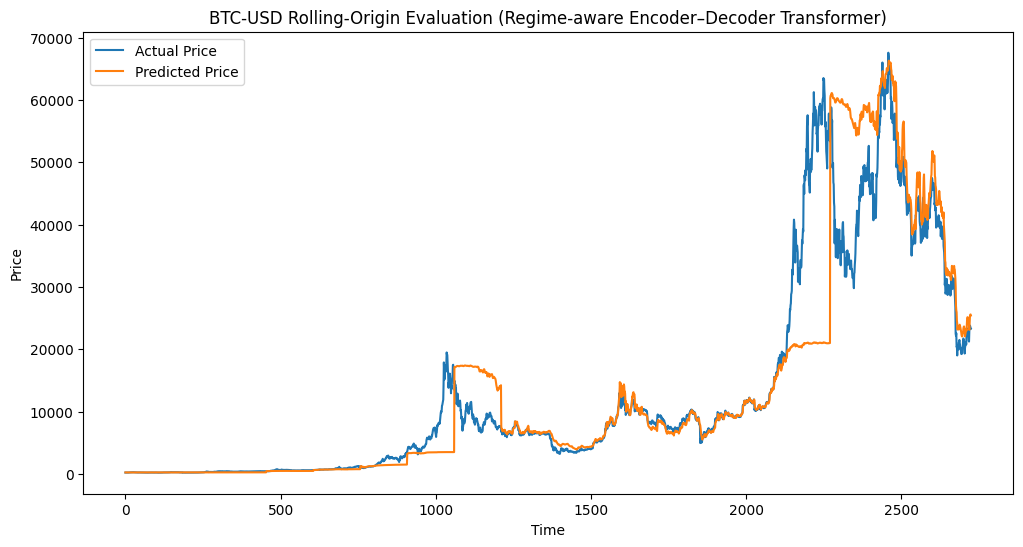}
    \caption{BTC-USD Rolling-Origin Evaluation using the Regime-aware Encoder--Decoder Transformer (Architecture 1).}
    \label{fig:btc_arch1}
\end{figure}

\begin{figure}[H]
    \centering
    \includegraphics[width=0.95\linewidth]{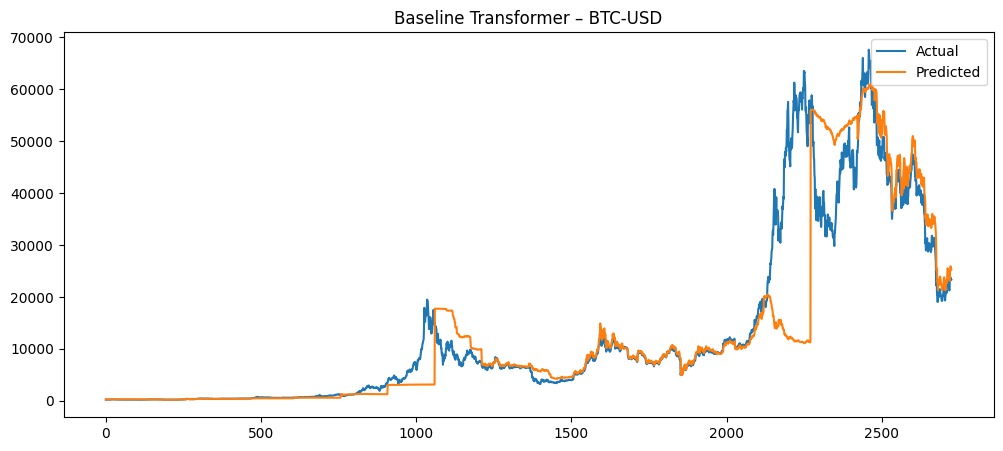}
    \caption{BTC-USD Rolling-Origin Evaluation using the Baseline Transformer.}
    \label{fig:btc_baseline}
\end{figure}

\begin{figure}[H]
    \centering
    \includegraphics[width=0.95\linewidth]{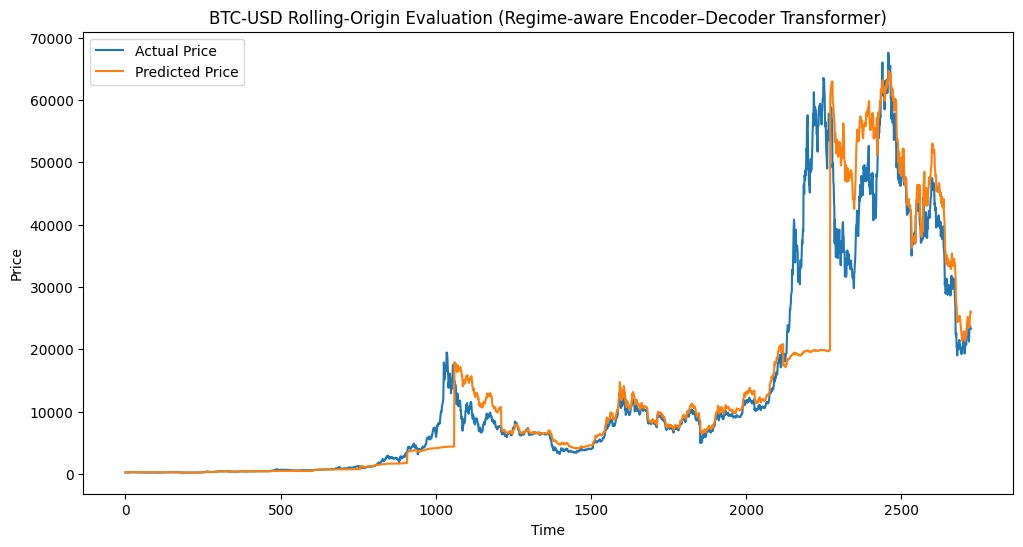}
    \caption{BTC-USD Rolling-Origin Evaluation using the Regime-aware Transformer (Architecture 0).}
    \label{fig:btc_arch0}
\end{figure}% \FloatBarrier

\vspace*{4cm}

\subsection{Visual Comparison of Key Metrics}

\begin{figure}[H]
    \centering
    \includegraphics[width=0.95\linewidth]{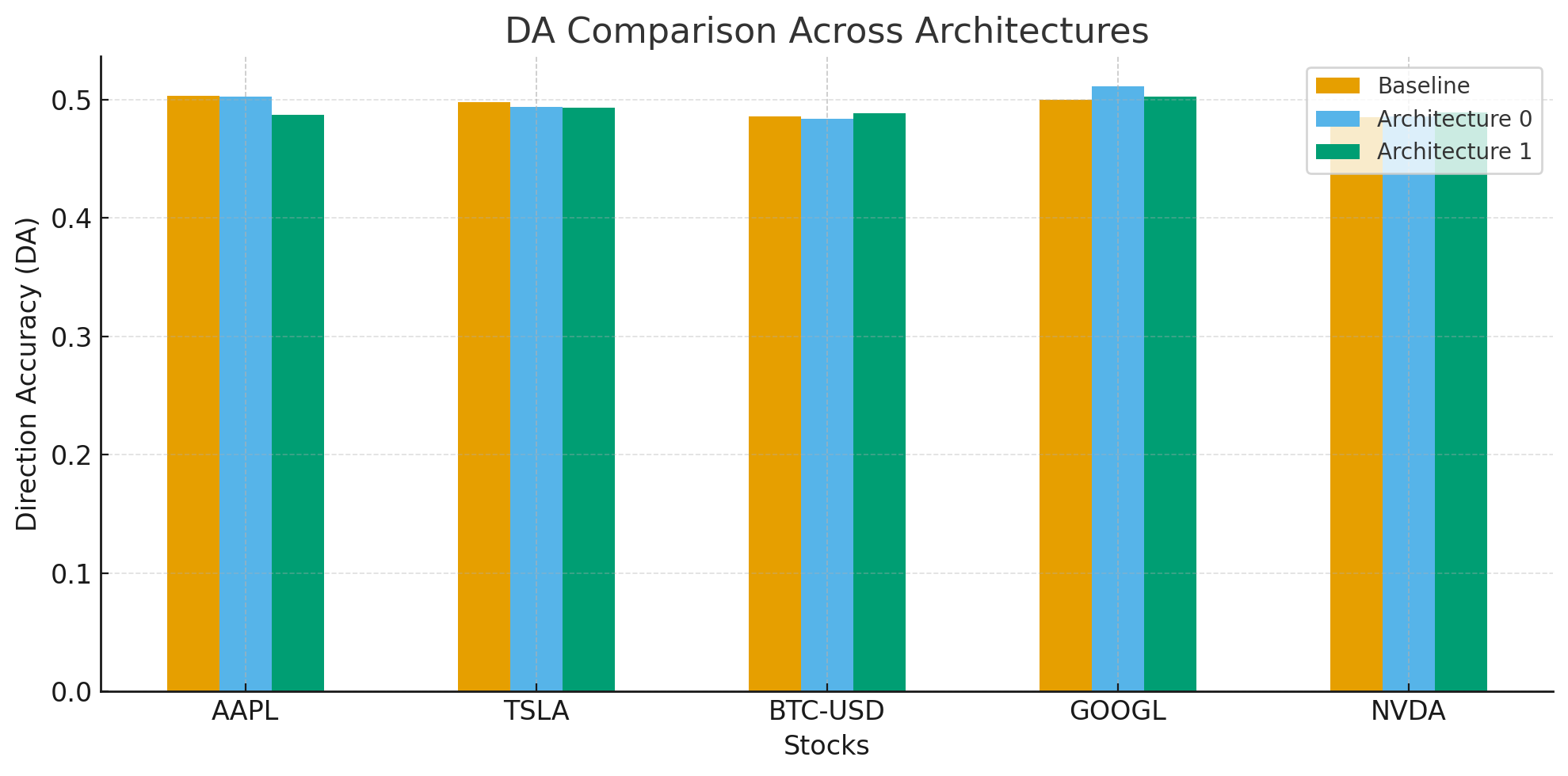}
    \caption{Directional accuracy comparison across all architectures.}
    \label{fig:da_plot}
\end{figure}

\begin{figure}[H]
    \centering
    \includegraphics[width=0.95\linewidth]{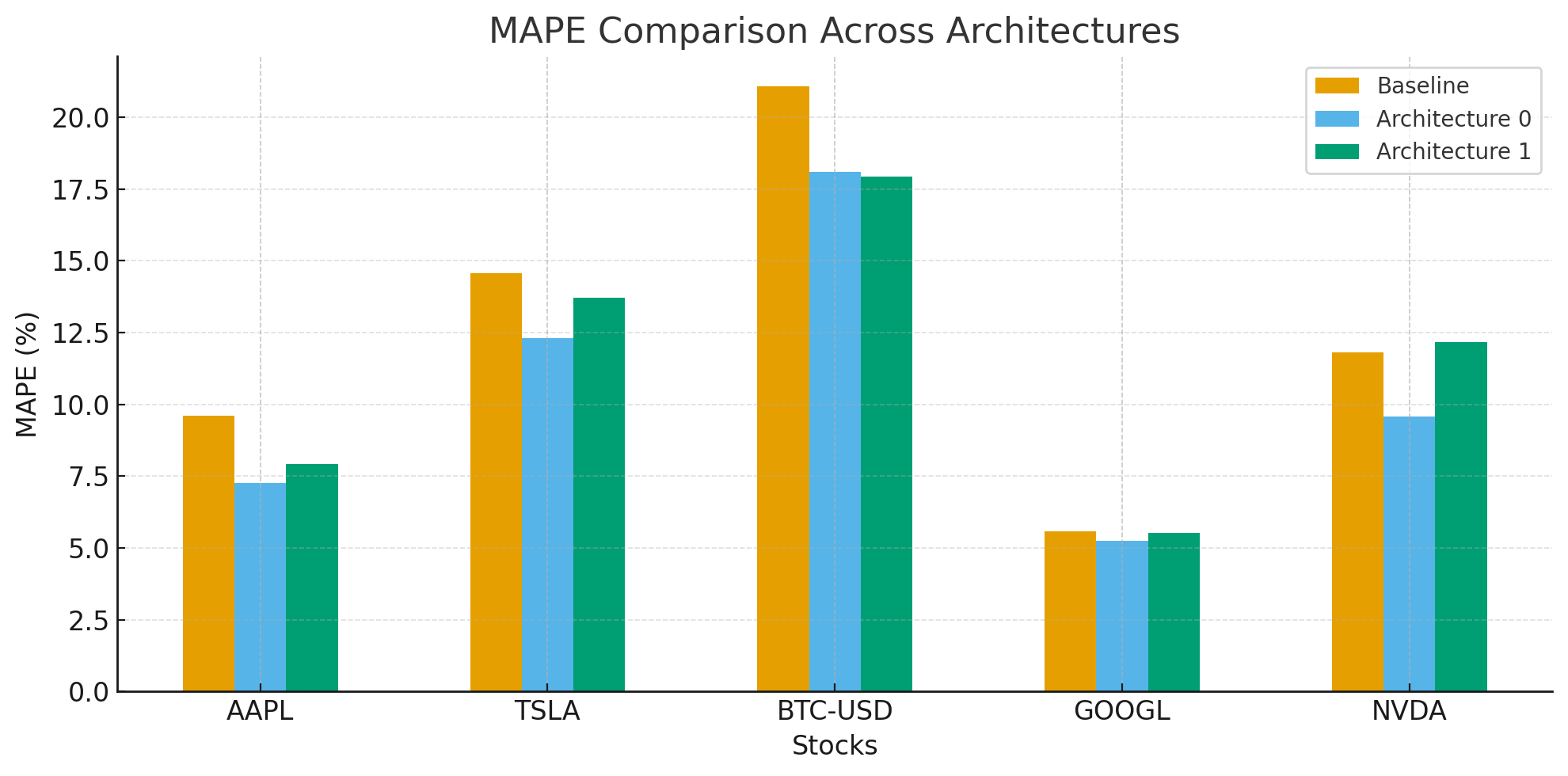}
    \caption{MAPE comparison across all architectures.}
    \label{fig:mape_plot}
\end{figure}

To provide additional interpretability, Figures~\ref{fig:da_plot} and 
\ref{fig:mape_plot} visualize aggregated results across all stocks for 
Directional Accuracy and MAPE, respectively.  
These two metrics summarize model performance from complementary perspectives:
trend prediction (DA) and numeric precision (MAPE).

% \twocolumn 

\subsection{Discussion}

Across nearly all stocks, the introduction of regime-awareness improves model 
robustness and forecasting consistency.  
Architecture~0 typically yields lower MAPE and higher correlation scores 
compared to the baseline, indicating that enriching embeddings with structural 
market context helps the model interpret the input sequence more effectively.

Architecture~1 further improves Directional Accuracy, suggesting that modulating
attention weights based on the current regime helps the model adapt its
effective receptive field.  
By altering the attention patterns through regime-derived biases, the model
learns to focus more on short-term dependencies during high-volatility periods
and to extend its temporal horizon when trends are persistent.

Overall, regime-aware mechanisms provide consistent improvements across
multiple asset classes and evaluation folds, demonstrating the benefit of
context-adaptive attention in financial forecasting models.

\section{Conclusion}

This work investigated how regime-aware representations can improve the ability
of Transformer models to forecast financial time series. Standard Transformers
operate with a fixed attention structure, which makes them less suited for
markets that frequently shift between high volatility, trending, and
mean-reverting regimes. By introducing regime-driven embeddings, we allow the
model to adapt its effective attention window and interpret historical prices in
the context of prevailing market conditions.

Both proposed architectures—one modifying input embeddings and the other
modulating attention logits—show consistent improvements over the baseline
Transformer across multiple assets and evaluation metrics. These results suggest
that incorporating structural market signals provides a simple yet effective way
to make deep sequence models more robust to non-stationary financial behavior.

\section{Acknowledgment}

This project was carried out under the guidance of Prof.~Amit Sethi as part of
EE782: Advanced Topics in Machine Learning at IIT Bombay. We are grateful for
his suggestions and discussions throughout the development of the models and
experiments, which significantly helped refine the ideas presented in this
work.

\bibliographystyle{IEEEtran}
\bibliography{citations}

\end{document}